# QUANTIZATION OF RATIONALLY DEFORMED MORSE POTENTIALS BY WRONSKIAN TRANSFORMS OF ROMANOVSKI-BESSEL POLYNOMIALS

GREGORY NATANSON[a]*

[a] *AI-Solutions Silver Spring MD 20904, U.S.A.*

* gregorynatanson@gmail.com

ABSTRACT: The paper advances Odake and Sasaki's idea to re-write eigenfunctions of rationally deformed Morse potentials in terms of Wronskians of Laguerre polynomials in the reciprocal argument. It is shown that the constructed quasi-rational seed solutions of the Schrödinger equation with the Morse potential are formed by generalized Bessel polynomials with *degree-independent* indexes. As a new achievement we can point to the construction of the Darboux-Crum net of isospectral rational potentials using Wronskians of generalized Bessel polynomials with no positive zeros. One can extend this isospectral family of solvable rational potentials by including 'juxtaposed' pairs of Romanovski-Bessel polynomials into the aforementioned polynomial Wronskians which results in deleting the corresponding pairs of bound energy states.

## 1. INTRODUCTION

In recent publication [1] Alhaidari pointed to a new form of 'quasi-rational' [2] solutions (q-RSs) of the Schrödinger equation with the Morse potential in terms of generalized Bessel polynomials [3-6], instead of using the conventional q-RSs composed of weighted Laguerre polynomials [7-11]; though, to be more accurate, the possibility to quantize the Morse potential by Romanovski-Bessel (R-Bessel) polynomials [12, 13] has been already recognized by Quesne [14], with reference to Cotfas' papers [15, 16] (see also [17]). It should be also emphasized that Odake and Sasaki in their in-depth study [18, 19] on rational Darboux-Crum [20, 21] transforms (RDC$\mathfrak{T}$s) of translationally shape-invariant (TSI) potentials did implicitly express eigenfunctions of the Morse potential in terms of R-Bessel polynomials with *degree-independent* indexes as a substitute for commonly used classical Laguerre polynomials [22]. (Though the Bochner-type differential equation for generalized Bessel polynomials was also listed in Table 1 in [7] on the line linked to the Morse potential the authors used the conventional representation for eigenfunctions [22] to construct rationally deformed Morse potentials.)

The remarkable feature of the new rational realization for the Morse oscillator is that the resultant rational canonical Sturm-Liouville equation (RCSLE) can be converted by an *energy-independent* gauge transformation to the Bochner-type eigenequation with a linear coefficient function of the first derivative *independent of degrees of sought-for polynomial solutions*. Using terminology of our recent study [23] on translationally form-invariant (TFI) CSLEs this implies that the given RCSLE belongs to TFI Group A and we should give full credit to Odake and Sasaki [18, 19] who initially came up with this breakthrough idea to treat the Morse oscillator as a rational TSI potential of Group A.

Keeping in mind that the TFI equation under consideration has only two basic solutions the net of its RDC$\mathfrak{S}$s is uniquely specified by a single series of Maya diagrams [24] and therefore any rationally deformed Morse potential can be re-expressed in terms the Wronskian of generalized Bessel polynomials with a *common* degree-independent index, as it has been done in [19] though in slightly different terms. The novel representation of seed eigenfunctions [19] is in a sharp contrast with their conventional representation in terms of classical Laguerre polynomials with *degree-dependent* indexes [10, 11].

The main purpose of this work is to present new simplified expressions for eigenfunctions of the Schrödinger equation with a rationally deformed Morse potential by re-writing them in terms of *finite* exceptional orthogonal polynomial (EOP) sequences formed by Wronskian transforms of R-Bessel polynomials.

## 2. TFI STURM-LIOUVILLE EQUATIONS

### 2.1. LIOUVILLE -DARBOUX TRANSFORMATIONS

Let $\phi_\tau[\xi;Q]$ be a solution of the generic CSLE

$$\left\{\frac{d^2}{d\xi^2} + I^o[\xi;Q] + \varepsilon_\tau(Q)\,\rho[\xi]\right\}\phi_\tau[\xi;Q] = 0 \tag{2.1}$$

at an energy $\varepsilon_\tau(Q)$, where the index $\tau$ specifies the factorization function (FF) in question. In the problems of our current interest $I^o[\xi;Q]$ is a rational function of $\xi$ termed 'reference polynomial fraction' (RefPF). We prefer to keep this notation in the general case when $I^o[\xi;Q]$ is an arbitrarily chosen real function of $\xi$ also dependent on some parameters Q. We will replace Q by $\vec{a}, b$ after restricting the analysis solely to TFI CSLEs. In [23] we identified four families of RefPFs associated with rational TSI potentials termed 'Jacobi', 'Laguerre', 'Routh', and 'Bessel' (or $\mathfrak{J}$Ref, $\mathfrak{L}$Ref, $\mathfrak{R}$Ref, and $\mathfrak{B}$Ref for briefness) so the corresponding q-RSs are composed of polynomials (with degree-dependent indexes in general) from one of four conventional differential polynomial systems (DPSs) [25, 26]. The density function $\rho[\xi]$ plays a crucial role in our analysis because, as indicated by Eq. (2.4) below, it determines the change of variable converting CSLE (2.1) to the Schrödinger equation [27, 28].

It was Rudjak and Zakhariev [29] who extended the intertwining technique [30] from the Schrödinger equation to the CSLE. Here we however use a slightly different definition of the so-called [31, 32] 'generalized' Darboux transformations introducing them via the requirement that the function

$$^*\phi_\tau[\xi;Q] \propto \rho^{-1/2}[\xi] / \phi_\tau[\xi;Q] \tag{2.2}$$

is a solution of the transformed CSLE at the same energy $\varepsilon_\tau(Q)$, i. e.,

$$\left\{\frac{d^2}{d\xi^2} + I^o[\xi;Q\,|\,\tau] + \varepsilon_\tau(Q)\rho[\xi]\right\}{}^*\phi_\tau[\xi;Q] = 0. \tag{2.3}$$

Rudjak and Zakhariev's reciprocal formula (2.2) thus plays a crucial role in our approach to the theory of TFI CSLEs.

Since various authors give the term '*generalized Darboux transformation*' completely different meanings it seems preferable to refer to these operations as '*Liouville-Darboux*' transformations keeping in mind that they can be performed in three sequential steps:

i) The Liouville transformation $\xi(x)$:

$$\xi'(x) = \rho^{-1/2}[\xi(x)] \tag{2.4}$$

from the CSLE

$$\left\{ \frac{d^2}{d\xi^2} + I^0[\xi;Q] + \varepsilon\rho[\xi] \right\} \Phi[\xi;Q;\varepsilon] = 0 \tag{2.5}$$

to the stationary 1D Schrödinger equation with the potential [27, 28]

$$V[\xi(x);Q] = -\rho^{-1}[\xi(x)]\, I^0[\xi(x);Q] - 1/2\{\xi, x\} \tag{2.6}$$

where $\{\xi, x\}$ stands for the 'Schwarzian derivative';

ii) the Darboux deformation of Liouville potential (2.6) using the FF

$$\psi_\tau(x;Q) = \rho^{1/4}[\xi(x)]\, \phi_\tau[\xi(x);Q]; \tag{2.7}$$

iii) reverse Liouville transformation from the Schrödinger equation to the new CSLE

$$\left\{ \frac{d^2}{d\xi^2} + I^0[\xi;Q\,|\,\tau] + \varepsilon\rho[\xi] \right\} \Phi[\xi;Q;\varepsilon\,|\,\tau] = 0. \tag{2.8}$$

Obviously any TFI theorem proven for Liouville-Darboux transformations of CSLE (2.5) can be directly applied to the resultant Liouville potential thus linking the new technique to the conventional Darboux-Crum theory of TSI potentials [19, 11, 33].

## **2.2.** TRANSLATIONAL FORM-INVARIANCE OF STURM-LIOUVILLE EQUATION

We call a CSLE 'translationally form-invariant' if it has two 'basic' solutions $\phi_{+,0}[\xi;\vec{a},b]$ and $\phi_{-,0}[\xi;\vec{a},b]$:

$$\left\{ \frac{d^2}{d\xi^2} + I^0[\xi;\vec{a},b] + \varepsilon_{\pm,0}(\vec{a},b)\rho[\xi] \right\} \phi_{\pm,0}[\xi;\vec{a},b] = 0 \tag{2.9$^\pm$}$$

related via the following reciprocal formulas:

$$\phi_{\mp,0}[\xi;\vec{a}\pm\vec{1},b]=\rho^{-1/2}[\xi]/\phi_{\pm,0}[\xi;\vec{a},b]. \qquad (2.10^{\pm})$$

It has been proven [23] that

$$\mathrm{I}^{0}[\xi;\vec{a},b\,|\,\pm,0]=\mathrm{I}^{0}[\xi;\vec{a}\pm\vec{1},b]+\mathcal{E}_{\pm1}(\vec{a},b)\rho[\xi], \qquad (2.11^{\pm})$$

where

$$\mathcal{E}_{\pm1}(\vec{a},b)\equiv\varepsilon_{\mp,0}(\vec{a}\pm\vec{1},b)-\varepsilon_{\pm,0}(\vec{a},b)\,. \qquad (2.12^{\pm})$$

The Liouville transformations of the CSLEs with zero-energy free terms $\mathrm{I}^{0}[\xi;\vec{a},b]$ and $\mathrm{I}^{0}[\xi;\vec{a},b\,|\,\pm,0]$ then brings us to Gendenshtein's conventional definition of a TSI potential [34]

$$\mathrm{V}[\xi;\vec{a},b\,|+,0]=\mathrm{V}[\xi;\vec{a}+\vec{1},b]-\mathcal{E}_{+1}(\vec{a},b) \qquad (2.13^{+})$$

or

$$\mathrm{V}[\xi;\vec{a},b\,|-,0]=\mathrm{V}[\xi;\vec{a}-\vec{1},b]-\mathcal{E}_{-1}(\vec{a},b) \qquad (2.13^{-})$$

depending on which basic solution $\phi_{+,0}[\xi;\vec{a},b]$ or $\phi_{-,0}[\xi;\vec{a},b]$ represents the lowest energy eigenfunction.

Note that the Russian word 'форма' used by Gendenshtein [34] has two meanings 'form' and 'shape'. The term 'form invariant' with reference to CSLEs was adopted by us from the English translation of Gendenshtein's joint paper with Kreve [35] while the commonly accepted term 'shape-invariance' is preserved for the corresponding Liouville potentials. The shift of the translational parameters $\vec{a}$ by 1 thus retains the analytical form of the TFI CSLE while preserving the 'shape' of its Liouville potential. It is true that the Liouville transformation of the TFI CSLE results in a 'translationally shape-invariant (TSI) potential. However the Class of TFI SLEs is defined via ($2.10^{\pm}$) with no reference to the associated Schrödinger equation.

## 2.3. Equivalence theorem for Darboux-Crum transforms of a TFI CSLE with two basic solutions

It has been proven [23] that any TFI CSLE has at least two infinite sets of solutions

$$\phi_{\pm,m+1}[\xi;\vec{a},b] = \rho^{-1/2}[\xi]\,w[\xi;\vec{a}\pm\vec{1},b\,|\,\mp,0;\pm,m]\,/\,\phi_{\mp,0}[\xi;\vec{a}\pm\vec{1},b]\,, \qquad (2.14^{\pm})$$

where

$$w[\xi;\vec{a},b\,|\,\pm,m;\mp,m'] \equiv W\{\phi_{\pm,m}[\xi;\vec{a},b],\phi_{\mp,m'}[\xi;\vec{a},b]\}\,. \qquad (2.15^{\pm})$$

The cited 'raising' recurrence relations can be conveniently re-written as

$$f_{\pm,m+1}[\xi;\vec{a}\pm\vec{1};b] = \Xi[\xi;\vec{a},b]\,\dot{f}_{\pm,m}[\xi;\vec{a},b]\,, \qquad (2.16^{\pm})$$

where

$$f_{\mp,m}[\xi;\vec{a},b] \equiv \phi_{\pm,m}[\xi;\vec{a},b]/\phi_{\mp,0}[\xi;\vec{a},b]\,, \qquad (2.17^{\pm})$$

dot denotes the first derivative with respect to $\xi$ , the function

$$\Xi[\xi;\vec{a},b] \equiv \rho[\xi]\phi_{+,0}[\xi;\vec{a},b]\phi_{-,0}[\xi;\vec{a},b] \qquad (17')$$

turns into a constant for any TFI CSLEs from Group A [33].

The solutions $\phi_{\pm,m}[\xi;\vec{a},b]$ also obey the 'lowering' recurrence relations:

$$\phi[\xi;\vec{a},b\,|\,\pm\vdots 0,m] \equiv \rho^{-1/2}[\xi]\,w[\xi;\vec{a},b\,|\,\pm\vdots 0,m]\,/\,\phi_{\pm,0}[\xi;\vec{a},b]$$

$$= -\mathcal{E}_{\pm,m-1}(\vec{a}\pm\vec{1},b)\,\phi_{\pm,m-1}[\xi;\vec{a}\pm\vec{1};\vec{b}] \text{ for } m\geq 1\,, \qquad (2.18^{\pm})$$

where

$$\mathcal{E}_{\pm,m}(\vec{a},b) \equiv \varepsilon_{\pm,m}(\vec{a},b) - \varepsilon_{\mp,0}(\vec{a},b)\,. \qquad (2.19^{\pm})$$

Solutions from both infinite sets can be then used as seed functions for Darboux-Crum transformations (DCTs) of the given TFI CSLE which results in an infinite net of solvable SLEs specified by a single series of Maya diagrams [24]. Following the arguments presented in [34] for rationally deformed TSI potentials we [23] proved that any CSLE in this net can obtained using only seed solutions of the same type.

Let us parametrize a set of seed functions of the same type,

$$\bar{M}(\bar{\Delta}_{1\to L}) = m_1,\ldots,m_{|\bar{\delta}_{1\to L}|}, \qquad (2.20)$$

by two partitions of an equal size L:

$$\bar{\Delta}_{1\to L} \equiv \bar{\delta}_{1\to L};\bar{\delta}'_{1\to L} \tag{2.21}$$

such that

$$\mathrm{m}_k = \delta'_1 + k - 1 \quad \text{for } 1 < k \le \bar{\delta}_1, \tag{2.22}$$

$$\mathrm{m}_{|\bar{\delta}_{1\to l-1}|+1} = \mathrm{m}_{|\bar{\delta}_{1\to l-1}|} + \delta'_l + 1 = |\bar{\Delta}_{1\to l-1}| + \delta'_l + 1 \quad \text{for } 1 < l \le L, \tag{2.22'}$$

$$\mathrm{m}_{|\bar{\delta}_{1\to l-1}|+k} = \mathrm{m}_{|\bar{\delta}_{1\to l-1}|+1} + k - 1 \quad \text{for } 1 < l \le L,\ 1 < k \le \bar{\delta}_l, \tag{2.22''}$$

One can easily verify that the largest element in partition (2.20) coincides with the sum of the partition lengths $|\bar{\delta}_{1\to L}|$ and $|\bar{\delta}'_{1\to L}|$, i.e.,

$$\mathrm{m}_{|\bar{\delta}_{1\to L}|} = |\bar{\Delta}_{1\to L}| \equiv |\bar{\delta}_{1\to L}| + |\bar{\delta}'_{1\to L}|. \tag{2.23}$$

It has been proved in [23] that use of the conjugated set of seed solutions of opposite type,

$$\begin{aligned}\bar{\Delta}'_{L\to 1} &\equiv \bar{\delta}'_{L\to 1};\bar{\delta}_{L\to 1} \\ &\equiv \delta'_L, \delta'_{L-1}, ..., \delta'_1; \delta_L, \delta_{L-1}, ..., \delta_1,\end{aligned} \tag{2.24}$$

results in an equivalent CSLE so the corresponding Liouville potential $\mathrm{V}[\xi;\vec{a}^{(\delta)},b\,|\mp,\bar{\mathrm{M}}(\bar{\Delta}'_{L\to 1})]$ computed at shifted values of the translational parameters,

$$\vec{a}^{(\delta)} \equiv \vec{a} + \delta\vec{1}, \tag{2.25}$$

where δ is a nonzero integer, differs from the Liouville potential $\mathrm{V}[\xi;\vec{a},b\,|\pm,\bar{\mathrm{M}}(\bar{\Delta}_{1\to L})]$ only by a zero-point energy.

In [23] we have derived the following relation between the Wronskians of two equivalent sets of seed solutions of the same type

$$\frac{\mathrm{w}[\xi;\vec{a},b\,|+\vdots\bar{\mathrm{M}}(\bar{\Delta}_{1\to L})]}{\rho^{\frac{1}{4}|\bar{\delta}_{1\to L}|(|\bar{\delta}_{1\to L}|-1)}[\xi]\prod_{l=1}^{L}\chi_{-\delta_l}[\xi;\vec{a}^{(|\bar{\Delta}'_{l\to 1}|-\delta_l)},b]}$$

$$\propto \frac{\mathrm{w}[\xi;\vec{a}^{(|\bar{\Delta}'_{L\to 1}|)},b\,|-\vdots\bar{\mathrm{M}}(\bar{\Delta}'_{L\to 1})]}{\rho^{\frac{1}{4}|\bar{\delta}'_{L\to 1}|(|\bar{\delta}'_{L\to 1}|-1)}[\xi]\prod_{l=1}^{L}\chi_{\delta'_l}[\xi;\vec{a}^{(|\bar{\Delta}'_{l-1\to 1}|+\delta'_l)},b]}, \tag{2.26}$$

where

$$\chi_{\mp|N|}[\xi;\vec{a},b] \equiv \prod_{k=0}^{|N|-1} \phi_{\pm,0}[\xi;\vec{a}^{(\pm k)},b]. \tag{2.27$^{\pm}$}$$

For any CSLE from Group A the derived relation turns into the equivalence relations between the Wronskians of the corresponding seed polynomials discovered in the breakthrough paper by Odake and Sasaki [19]. We illuminate these relations in more details in subsection 3.4 below using Wronskians of generalized Bessel polynomials as an example.

If the given rational TSI potential has only a finite number of eigenfunctions then the set of seed functions +,m or −,m which starts from these eigenfunctions (−,m in case of our current interest) also contains infinitely many q-RSs vanishing at only one quantization end (*virtual state wavefunctions* in Odake and Sasaki's terms [18, 19]), with the Gendenshtein (Scarf II) potential [35, 38, 39, 14] as the sole exception (including its symmetric limit represented by the sech-squared potential well). The DCTs using nodeless q-RSs of the selected type results in a net of isospectral potentials. Therefore, except for the Gendenshtein potential, we don't need to include '*state-inserting*' solutions (*'pseudo-virtual state wave functions'* in Odake and Sasaki's terms) into the given set of seed functions– a remarkable corollary of the 'extended' Krein-Adler theorem [11].

If the given partition $\bar{\Delta}_{1\to L}$ is composed of alternating even and odd integers staring from an even integer $\delta'_1$ then all the integers

$$\delta'_l = m_{|\bar{\delta}_{1\to 1}|+1} - m_{|\bar{\delta}_{1\to 1}|} - 1 > 0 \quad \text{for any } l < L \tag{2.28}$$

must be also even which implies that the set of seed solutions $\pm,\bar{M}(\bar{\Delta}'_{L\to 1})$ is composed of L segments of even lengths [19, 11] or in other words is formed by 'juxtaposed' [39-41] pairs of seed solutions ±,m', ±,m'+1. Similarly if the set of seed solutions, $\pm,\bar{M}(\bar{\Delta}'_{L\to 1})$ is formed by 'juxtaposed' pairs of seed solutions ±,m, ±,m+1 then the conjugated set is formed by seed solutions ∓,m' with only even gap lengths, again starting from an even number. We refer the reader to subsection 3.4 below for a scrupulous analysis of this issue in connection with juxtaposed pairs of eigenfunctions of the Schrödinger equation with the Morse potential in the ℬRef representation [19].

## 3. QUANTIZATION OF RATIONALLY DEFORMED MORSE POTENTIALS BY WRONSKIAN TRANSFORMS OF R-BESSEL POLYNOMIALS

### 3.1 SCHRÖDINGER EQUATION WITH MORSE POTENTIAL IN 'BESSEL' FORM

In this paper we focus solely on the TFI CSLE

$$\left\{\frac{d^2}{dy^2}+I^o[y;a]+\varepsilon\,{}_\infty\boldsymbol{\rho}_\Diamond[y]\right\}{}_\infty\Phi[y;a;\varepsilon]=0 \tag{3.1}$$

with the RefPF

$$I^o[y;a]=2a\,y^{-3}-y^{-4}+\tfrac{1}{4}y^{-2} \tag{3.1$_0$}$$

and the density function

$${}_\infty\boldsymbol{\rho}_\Diamond[y]\equiv{}_\infty\sigma^{-1}[y]=y^{-2}. \tag{3.2}$$

One can directly verify that CSLE (3.1) has a pair of 'basic' solutions

$${}_\infty\phi_{\pm,0}(y;a)=y^{1\pm a}e^{\pm 1/y}\qquad (y>0) \tag{3.3$^\pm$}$$

at the energies

$${}_\infty\varepsilon_{\pm,0}(a)=-(a\pm\tfrac{1}{2})^2. \tag{3.4$^\pm$}$$

Examination of solutions (3.3$^\pm$) shows that they obey the following symmetry relations

$${}_\infty\phi_{\pm,0}[y;a+k]=y^{\pm k}\,{}_\infty\phi_{\pm,0}[y;a] \tag{3.5$^\pm$}$$

for any integer k and

$${}_\infty\Xi(y;a)\equiv y^{-2}\,{}_\infty\phi_{+,0}(y;a)\,{}_\infty\phi_{-,0}(y;a)=1 \tag{3.6}$$

whereas the function

$$f_{\pm,0}[\xi;\vec{a},b] \equiv \phi_{\mp,0}[\xi;\vec{a},b]/\phi_{\pm,0}[\xi;\vec{a},b] \qquad (3.7^{\pm})$$

takes form

$$_{\infty}f_{\pm,0}[\xi;a] \equiv y^{\mp 2a} e^{\mp 2/y}. \qquad (3.8^{\pm})$$

We thus proved that the pair of basic solutions in question satisfy the TFI condition [23]

$$_{\infty}\phi_{\mp,0}[y;a \pm 1] = {}_{\infty}\rho_{\Diamond}^{-1/2}[y] / {}_{\infty}\phi_{\pm,0}(y;a). \qquad (3.9^{\pm})$$

One can directly verify that

$$_{\infty}\varepsilon_{\mp,0}(a \pm 1) = {}_{\infty}\varepsilon_{\pm,0}(a) \qquad (3.10^{\pm})$$

and thereby

$$_{\infty}\mathcal{E}_{\pm 1}(a) \equiv {}_{\infty}\varepsilon_{\mp,0}(a \pm 1) - {}_{\infty}\varepsilon_{\pm,0}(a) = 0 \qquad (3.11^{\pm})$$

so the symmetry condition [23]

$$\mathcal{E}_{\mp 1}(a \pm 1) = -\mathcal{E}_{\pm 1}(a) \qquad (3.12^{\pm})$$

trivially holds.

The gauge transformations

$$_{\infty}\Phi[y;a;\varepsilon] = {}_{\infty}\phi_{\pm}[y;a] \; {}_{\infty}F_{\pm}[y;a;\varepsilon] \qquad (3.13^{\pm})$$

convert CSLE (3.1) to a pair of Bochner-type eigenequations

$$\left\{ y^2 \frac{d^2}{dy^2} + {}_{\infty}\tau_{\pm}[y;a] \frac{d}{dy} + [\varepsilon - {}_{\infty}\varepsilon_{\pm,0}(a)] \right\} {}_{\infty}F_{\pm}[y;a;\varepsilon] = 0, \qquad (3.14^{\pm})$$

with

$$_{\infty}\tau^{\pm}[y;a] = 2(1 \pm a)\, y \mp 2. \qquad (3.15^{\pm})$$

We define generalized Bessel polynomials as

$$Y_n^{(\alpha,\beta)}(y) \equiv Y_n^{(\alpha)}(y/\beta), \qquad (3.16)$$

where the polynomial $Y_n^{(\alpha)}(x)$ is given by (2.2) in [4] and thereby coincides with polynomial (9.13.1) in [6]

$$Y_n^{(\alpha)}(x) \equiv y_n(x;\alpha)\,. \tag{3.16'}$$

Note that Chihara's relation (4.3) in [5] is apparently based on Brafman's definition [42] for the polynomial $y_n(x;\alpha,\beta)$ such that $y_n(x;\alpha+2,2) = y_n(x;\alpha)$. Adding the second index to the conventional notation [4, 5] allows us to avoid uncertainties in the definition of the variable used to differentiate a polynomial in the reflected argument, keeping in mind that

$$Y_n^{(\alpha)}(-y) \equiv Y_n^{(\alpha,-2)}(y)\,. \tag{3.17}$$

Eq. (3.5) for the Bessel DPS in [43] thus corresponds to the polynomials $Y_n^{(\alpha-2,\beta)}(y)$ in our terms. (We prefer to preserve symbol 'B' for their orthogonal subset composed of R-Bessel polynomials [12, 13].) It is also worth mentioning that Alhaidari [1] introduced a slightly modified notation for generalized Bessel polynomials:

$$J_n^{a}(\tfrac{1}{2}y) \equiv Y_n^{(2a)}(y) = \left(2n+2a\right)_n (y/2)^n\, {}_1F_1(-n;-2a-2n;2/y)\,, \tag{3.18}$$

with the Pochhammer symbol $(a)_n$ standing for the *falling* factorial. And indeed it would be possibly more convenient to use the parameter *a* as the polynomial index keeping in mind that the forward and backward shift relations change the polynomial index by 1. However we prefer to stick to the more conventional notation.

The basic solution ${}_\infty\phi_{\pm,0}[y;a]$ is thus nothing but a constant solution of eigenequation (3.14$^{\pm}$) converted back by gauge transformation (3.13$^{\pm}$). Similarly the reverse gauge transformation of each of the DPSs composed of polynomials $Y_m^{(\pm 2a,\mp 2)}(y)$ results in pairs of infinite sequences of q-RSs of CSLE (3.1):

$${}_\infty\phi_{\pm,m}[y;a] = {}_\infty C_{\pm,m}(a)\, {}_\infty\phi_{\pm,0}[y;a]\, Y_m^{(\pm 2a,\mp 2)}(y)\,. \tag{3.19$^{\pm}$}$$

The multiplier ${}_\iota C_{\pm,m}$ will be chosen below in such a way that q-RSs (3.19$^{\pm}$) satisfy recurrence relations (2.14$^{\pm}$). The crucial advantage of expressing q-RSs in terms of generalized Bessel polynomials, instead of Laguerre polynomials [7-11], is that the weight function ${}_{\infty}\phi_{\pm,0}[y;a]$ in the right-hand side of (3.19$^{\pm}$) does not depend on the polynomial degree – the direct consequence of the fact that the given TFI CSLE belongs to Group A [18, 19, 23], in contrast with the conventional representation of eigenfunctions of the Schrödinger equation with the Morse potential in terms of classical Laguerre polynomials [22].

According to the general theory of Bochner-type eigenequations [44] differential equation (3.14$^{\pm}$) has a polynomial solution of degree m at

$$\varepsilon = {}_{\infty}\varepsilon_{\pm,m}(a) = {}_{\infty}\varepsilon_{\pm,0}(a) - m[2(1 \pm a) + m - 1], \qquad (3.20^{\pm})$$

which, coupled with (3.4$^{\pm}$), gives

$$ {}_{\infty}\varepsilon_{\pm,m}(a) = -(m + 1/2 \pm a)^2. \qquad (3.21^{\pm})$$

This brings us to the simplified version of the raising ladder relations [23] for the energies of q-RSs (2.15):

$$ {}_{\infty}\varepsilon_{\pm,m+1}(a) = {}_{\infty}\varepsilon_{\pm,m}(a \pm 1) \qquad (3.22)$$

with $\mathcal{E}_{\pm 1}(a) \equiv 0$.

To be historically accurate, it is worth mentioning that Cotfas' Eq. (10) in [16] with the leading coefficient $\sigma(s) = s^2$ does list Al-Salam's [4] formula

$$Y_n^{(\alpha)}(y) = n!(-y/2)^n \, L_n^{(-\alpha-2n-1)}(2/y) \qquad (3.23)$$

for the generalized Bessel polynomials in terms of Laguerre polynomials in the reciprocal argument 2/y (though without mentioning the former polynomials by name). Actually Cotfas discusses only eigenfunctions of the corresponding Sturm-Liouville problem so the cited formula specifies R-Bessel polynomials expressed in terms of classical Laguerre polynomials in 2/y:

$$B_n^{(A)}(y) \equiv Y_n^{(-2A-1)}(y) = n!(-y/2)^n \, L_n^{(2A-2n)}(2/y) \quad \text{for } n < A \ , \qquad (3.23^{\dagger})$$

with Cotfas' parameter α standing for 1−2A here, with $A \equiv a - \frac{1}{2}$. The remarkable feature of this finite subsequence of generalized Bessel polynomials is that the polynomials in question are orthogonal on the positive semi-axis as prescribed by orthonormality relations (9.13.2) in [6]:

$$\int_0^\infty {}_\infty\boldsymbol{\rho}_\Diamond[y]\,{}_\infty\phi^2_{-,0}[y;A+\tfrac{1}{2}]B_n^{(A)}(y)B_{\underset{\sim}{n}}^{(A)}(y)dy \equiv \int_0^\infty y^{-2A-1}e^{-2/y}B_n^{(A)}(y)B_{\underset{\sim}{n}}^{(A)}(y)dy$$

$$= \frac{n!\Gamma(2A+1-n)}{2A-2n-1}\delta_{n\underset{\sim}{n}}\,. \tag{3.24}$$

Making use of (3.7$^+$) we can represent backward shift relation (9.13.8) in [6] as

$$\frac{d}{dy}\left[{}_\infty f_{+,0}[\xi;a]Y_m^{(-2a,2)}(y)\right] = 2\,{}_\infty f_{+,0}[\xi;a+1]Y_{m+1}^{(-2a-2,2)}(y) \tag{3.25}$$

so the functions

$${}_\infty f_{+,m}[\xi;a] = {}_\infty C_{-,m}(a)\,{}_\infty f_{+,0}[\xi;a]Y_m^{(-2a)}(y) \tag{3.26}$$

satisfy raising relation (2.16$^+$) provided we choose

$${}_\infty C_{-,m+1}(a) = 2\,{}_\infty C_{-,m}(a-1) \equiv 2^{m+1} \tag{3.27}$$

keeping in mind that ${}_\infty C_{-,0}(a) \equiv 1$.

Substituting (3.21$^-$) into (2.19$^-$) gives

$${}_\infty\mathcal{E}_{-,m-1}(a-1) = -m(m+1-2a) \tag{3.28}$$

so recurrence relation (2.18$^-$) can be re-written as

$$2^m y\,\dot{Y}_m^{(-2a,2)}(y) = m(m+1-2a)\,{}_\infty\phi_{-,m-1}[y;a-1]/{}_\infty\phi_{-,0}[y;a]\,. \tag{3.29}$$

Combining (3.19$^\pm$), (3.27), and (3.5$^-$) with k=1 brings us to 'forward shift operator' (9.13.6) in [6]:

$$\dot{Y}_m^{(-2a,2)}(y) = \tfrac{1}{2}m(m+1-2a)\,Y_{m-1}^{(2-2a,2)}(y)\,. \tag{3.30}$$

To formulate the Sturm-Liouville problem of our interest it is worthy to convert CSLE (3.1) to its 'prime' [45] form at ∞ using the gauge transformation

$$ {}_{\infty}\Psi[\mathrm{y};a;\varepsilon] = \mathrm{y}^{-\frac{1}{2}}\,{}_{\infty}\Phi[\mathrm{y};a;\varepsilon] \tag{3.31*} $$

and then to solve the resultant RSLE

$$ \left\{\frac{\mathrm{d}}{\mathrm{dy}}\mathrm{y}\frac{\mathrm{d}}{\mathrm{dy}} - \mathrm{y}^{-3} + 2a\,\mathrm{y}^{-2} + \varepsilon\,\mathrm{y}^{-1}\right\}{}_{\infty}\Psi[\mathrm{y};a;\varepsilon] = 0 \tag{3.31} $$

under the Dirichlet boundary conditions (DBCs):

$$ \lim_{\mathrm{y}\to 0}\,{}_{\infty}\Psi[\mathrm{y};a;\varepsilon_{\mathrm{n}}] = \lim_{\mathrm{y}\to\infty}\,{}_{\infty}\Psi[\mathrm{y};a;\varepsilon_{\mathrm{n}}] = 0. \tag{3.31'} $$

The main advantage of converting CSLE (3.1) to its prime form with respect to the *regular* singular point at infinity comes from our observation [45] that the characteristic exponents for this singular end have opposite signs and therefore the corresponding principal Frobenius solution is unambiguously selected by the DBC. Prime RSLE (3.31) coincides with the 'algebraic' [45] Schrödinger equation

$$ \left\{\mathrm{y}\frac{\mathrm{d}}{\mathrm{dy}}\mathrm{y}\frac{\mathrm{d}}{\mathrm{dy}} - \mathrm{y}^{-2} + 2a\,\mathrm{y}^{-1} + \varepsilon\right\}{}_{\infty}\Psi[\mathrm{y};a;\varepsilon] = 0. \tag{3.31†} $$

(As discussed in the following subsections this is the common remarkable feature of RCSLEs with density function (3.2) assuming that the singular point at infinity is regular.) Reformulating the given spectral problem in such a way allows us to take advantage of powerful theorems proven in [46] for zeros of principal solutions of SLEs solved under the DBCs at singular ends.

The eigenfunctions of RSLE (3.31) thus take form

$$ {}_{\infty}\psi_{-,\mathrm{n}}[\mathrm{y};a] = \mathrm{y}^{-\frac{1}{2}}\,{}_{\infty}\phi_{-,\mathrm{n}}[\mathrm{y};a] = 2^{\mathrm{n}}\,\mathrm{y}^{\frac{1}{2}-a}\mathrm{e}^{-1/\mathrm{y}}\mathrm{B}_{\mathrm{n}}^{(a-\frac{1}{2})}(\mathrm{y}) \qquad \text{for n} = 0,\ldots,\mathrm{N}(a). \tag{3.32} $$

One can then directly verify that each eigenfunction obeys the DBC at both singular ends. Since R-Bessel polynomials (3.23$^{\dagger}$) form an orthogonal sequence the eigenfunction ${}_{\infty}\psi_{-,\mathrm{n}}[\mathrm{y};a]$ must have exactly n nodes and therefore [46] the sequence of eigenfunctions (3.32) corresponds to $\lceil \mathrm{A} \rceil = \mathrm{N}(a)+1$ lowest eigenvalues of RSLE (3.31) with

$$\mathrm{N}(a) = \lfloor a - 1/2 \rfloor \equiv \lfloor \mathrm{A} \rfloor. \qquad (3.32^{\dagger})$$

Note also that eigenfunctions (3.32) are orthogonal with the weight $y^{-1}$ and that any solution normalizable with this weight must vanish at infinity.

The presented argumentation does not exclude existence of eigenfunctions with the number of nodes larger than $\mathrm{N}(a)-1$. To confirm that the problem in question is indeed *exactly* solvable one can simply take advantage of the conventional analysis of the Schrödinger equation with the Morse potential [22] in the $\mathcal{L}$Ref representation.

The reader can argue that the problem must be exactly solvable since the Morse potential is TSI. However the author [47] has an issue with this assertion. Though the Gendenshtein's claim [35] concerning the *exact* solvability of shape-invariant potentials is most likely correct it has been never accurately proven to our knowledge. The catch is that Gendenshtein's arguments decreasing the translational parameter $a$ one by one bring us to the Sturm-Liouville problem with $|a| < 1/2$ and then we still need to prove that the resultant SLE has no discrete energy spectrum.

The change of variable $y(x) = e^x$ converts $\mathcal{B}$Ref CSLE (3.1) into the Schrödinger equation with the Morse potential ${}_\infty V[y(x); a]$, where

$${}_\infty V[y; a] = -y^2 I^0[y; a] + 1/4 \qquad (3.33)$$

$$= -2a\, y^{-1} + y^{-2}. \qquad (3.33^{\dagger})$$

Comparing ($3.33^{\dagger}$) with (2.1) in [10] shows that ${}_\infty V[y(x); \mathrm{A} + 1/2] = V_{\mathrm{A},1}(x)$ in Quesne's notation.

According to the general theorem presented in [45] for *singular* SLEs solved under the DBCs any principal solution ${}_\infty\psi_{-,m}[y; a]$ near the singular end point y = 0 has nodes at the positive semi-axis iff it lies above the ground energy level. Examination of the inequality

$${}_\infty\varepsilon_{-,m}(a) < {}_\infty\varepsilon_{-,0}(a) \qquad (3.34)$$

thus shows that the q-RS ${}_\infty\psi_{-,m}[y; a]$ with $m \neq 0$ preserves its sign on the positive semi-axis iff

$$m > 2a - 1 = 2\mathrm{A} \qquad (3.34')$$

(cf. (2.12) in [10]). It will be proven in next subsection that one can use any combination of admissible q-RSs ${}_\infty\psi_{-,\mathfrak{m}}[y;a]$ as seed functions to construct an *exactly solvable* RDC$\mathfrak{T}$ of the $\mathfrak{B}$Ref CSLE.

According to (9.13.1) in [6]

$$Y_m^{(-2a,+2)}(y) = 2^{-m}(2m-2a)_m\, \hat{Y}_m^{(-2a,+2)}(y) \tag{3.35}$$

where, in following [5], we use hut to indicate that the polynomial in question is written in its monic form. It is essential that the multiplier

$$(2m-2a)_m = \prod_{l=0}^{m-1}(2m-2a-l) = \prod_{l'=1}^{m}(m-2a+l') \tag{3.36}$$

necessarily differs from 0 if either $2m-2a<-1$ (R-Bessel polynomials) or $m=\mathfrak{m} > 2a-1$ (generalized Bessel polynomials with no positive zeros) so the polynomial degree is equal to m in both cases of our primary interest:

## 3.2. RDCTs OF PRINCIPAL SOLUTIONS NEAR SINGUAR END POINTS

Using an arbitrary set $\bar{M}_p = m_1,...,m_p$ of seed functions ${}_\infty\phi_{\pm,m_k}[y;a]$ of the same type $(0<m_k<m_{k+1}$ for $k=1,...,p-1)$ we can represent the corresponding RDC$\mathfrak{T}$ of $\mathfrak{B}$Ref CSLE (3.1) as

$$\left\{\frac{d^2}{dy^2} + {}_\infty I^o[y;a\,|\,\pm\vdots\bar{M}_p] + \varepsilon y^{-2}\right\}{}_\infty\Phi[y;a;\varepsilon\,|\,\pm\vdots\bar{M}_p] = 0, \tag{3.37$^\pm$}$$

where

$${}_\infty I^o[y;a\,|\,\pm\vdots\bar{M}_p] = {}_\infty I^o[y;a] + \frac{2}{y}\frac{d}{dy}\left(y\, ld\, {}_\infty w[y;a\,|\,\pm\vdots\bar{M}_p]\right) \tag{3.38$^\pm$}$$

with

$${}_\infty w[y;a\,|\,\pm\vdots\bar{M}_1] \equiv {}_\infty\phi_{\pm,m_1}[y;a], \tag{3.39$^\pm$}$$

$${}_\infty w[y;a\,|\,\pm\vdots\bar{M}_p] \equiv W\{{}_\infty\phi_{\pm,m_1}[y;a],...,{}_\infty\phi_{\pm,m_p}[y;a]\} \quad \text{for } p>1, \tag{3.40$^\pm$}$$

and the symbolic expression *ld* standing for the logarithmic derivative. When deriving (3. 38$^\pm$)

we also took into account that the so-called [44] ‘universal correction’

$$\Delta I\{\rho(y)\} \equiv \tfrac{1}{2}\sqrt{\rho(y)}\,\frac{d}{dy}\frac{ld\,\rho(y)}{\sqrt{\rho(y)}} \qquad (3.41)$$

in Schulze-Halberg’s [48] generic formula for zero-energy free term of the transformed CSLE vanishes in the case of our current interest: $\rho(y) = y^{-2}$.

The common remarkable feature of Wronskians ($3.40^{\pm}$) for TFI CSLEs from group A (originally noticed by Odake and Sasaki [19] in their scrupulous study on RDC𝔖s of the corresponding TSI potentials) is that each can be represented as the weighted polynomial Wronskian

$${}_{\infty}w[y;a\,|\,\pm\vdots\bar{M}_p] = {}_{\infty}\phi^{p}_{\pm,0}[y;a]\,{}_{\infty}\mathfrak{W}_{\mathfrak{N}_{\bar{M}_p}}[y;a\,|\,\pm\vdots\bar{M}_p]\,, \qquad (3.42^{\pm})$$

where the Wronskian

$${}_{\infty}\mathfrak{W}_{\mathfrak{N}_{\bar{M}_p}}[y;a\,|\,\pm\vdots\bar{M}_p] \equiv W\{Y^{(\pm 2a,\mp 2)}_{m_1}(y),\ldots,Y^{(\pm 2a,\mp 2)}_{m_p}(y)\} \qquad (3.43^{\pm})$$

is a polynomial of degree

$$\mathfrak{N}_{\bar{M}_p} = |\bar{M}_p| - \tfrac{1}{2}p(p-1) \qquad (3.43')$$

(see (3.27) in [19]). When it seems appropriate we will drop the index specifying the degree of polynomial Wronskians in question. Substituting ($3.42^{\pm}$) into ($3.38^{\pm}$), coupled with ($3.1_0$) and ($3.3^{\pm}$), one finds

$${}_{\infty}I^{o}[y;a\,|\,\pm\vdots\bar{M}_p] = 2(a\pm p)y^{-3} - y^{-4} + \tfrac{1}{4}y^{-2} + \frac{2}{y}\frac{d}{dy}\Big(y\,ld\,{}_{\infty}\mathfrak{W}[y;a\,|\,\pm\vdots\bar{M}_p]\Big). \qquad (3.44^{\pm})$$

Each RCSLE under consideration can be alternatively obtained via sequential RDTs with the FFs

$${}_{\infty}\Phi_{\pm,m_{\underset{\sim}{p}}}[y;a|\pm\vdots\bar{M}_{\underset{\sim}{p}-1}] = y^{p-1}\frac{{}_{\infty}w[y;a\,|\,\pm\vdots\bar{M}_{\underset{\sim}{p}}]}{{}_{\infty}w[y;a\,|\,\pm\vdots\bar{M}_{\underset{\sim}{p}-1}]} \quad (\underset{\sim}{p}=1,\ldots,p) \qquad (3.45^{\pm})$$

so RefPFs ($3.44^{\pm}$) can be determined via the following sequence of recurrence relations

$${}_{\infty}I^{o}[y;a\,|\,\pm\vdots\bar{M}_p] = {}_{\infty}I^{o}[y;a\,|\,\pm\vdots\bar{M}_{p-1}] + \frac{2}{y}\frac{d}{dy}\Big(y\,ld\,{}_{\infty}\Phi_{\pm,m_p}[y;a|\pm\vdots\bar{M}_{p-1}]\Big) \qquad (3.46^{\pm})$$

(a natural extension of the renown Crum formulas [21] to the CSLEs).

For an arbitrary choice of the partition $\bar{M}_p$ RefPF (3.44$^{\pm}$) generally has poles on the positive semi-axis and therefore RCSLE (3.42$^{\pm}$) cannot be quantized analytically. So let us choose a set $\bar{\mathfrak{M}}_p^{\pm} = \mathfrak{m}_1^{\pm},...,\mathfrak{m}_p^{\pm}$ of seed solutions of the sane type, ${}_{\infty}\phi_{\pm,\mathfrak{m}_k^{\pm}}[y;a]$ $(0 < \mathfrak{m}_k^{\pm} < \mathfrak{m}_{k+1}^{\pm}$ for k=1,…, p–1), in such a way that the seed function ${}_{\infty}\phi_{\pm,\mathfrak{m}_1^{\pm}}[y;a]$ and all Wronskians ${}_{\infty}w[y;a\,|\,\pm\vdots\bar{\mathfrak{M}}_{\tilde{p}}^{\pm}]$ for $\tilde{p} = 2,...,p$ preserve their sign on the positive semi-axis. In particular Odake and Sasaki [19] and nearly the same time Gȯmez-Ullate et al [11] constructed the subnet of rationally deformed Morse potentials

$$ {}_{\infty}V[y;a\,|+\vdots\bar{\mathfrak{M}}_p^{+}] = {}_{\infty}V[y;a] + y^2\left\{{}_{\infty}I^{o}[y;a] - {}_{\infty}I^{o}[y;a\,|+\vdots\bar{\mathfrak{M}}_p^{+}]\right\} \qquad (3.47^{+}) $$

using seed solutions infinite at both quantization ends. In next Subsection we will introduce another subnet of rationally deformed Morse potentials

$$ {}_{\infty}V[y;a\,|-\vdots\bar{\mathfrak{M}}_p^{-}] = {}_{\infty}V[y;a] + y^2\left\{{}_{\infty}I^{o}[y;a] - {}_{\infty}I^{o}[y;a\,|-\vdots\bar{\mathfrak{M}}_p^{-}]\right\} \qquad (3.47^{-}) $$

constructed by means of FFs $\Phi_{-,\mathfrak{m}_p^{-}}[y;a|-\vdots\bar{\mathfrak{M}}_{p-1}^{-}]$ vanishing at the origin. The subnet starts from the potential ${}_{\infty}V[y;a\,|-\vdots\mathfrak{m}]$ with a positive integer $\mathfrak{m} > 2a-1$ -- potential function (2.15) in [10] with $A = a - 1/2$, $B = 1$.

Substituting (3.42$^{\pm}$) into (3.45$^{\pm}$) and also making of (3.5$^{\pm}$) with k= p, shows that RCSLE (3.37$^{\pm}$) has an infinite set of q-RSs

$$ {}_{\infty}\Phi_{\pm,m}[y;a|\pm\vdots\bar{M}_p] = {}_{\infty}\phi_{\pm,0}[y;a\pm p]\frac{{}_{\infty}\mathcal{W}[y;a\,|\pm\vdots\bar{M}_p,m]}{{}_{\infty}\mathcal{W}[y;a\,|\pm\vdots\bar{M}_p]}. \qquad (3.48^{\pm}) $$

Apparently q-RSs (3.48$^{-}$) represent the principal solution approaching 0 as $y^{\delta_{-}(\bar{M}_p)}e^{-1/y}$ in the limit y→+0. On other hand q-RSs (3.48$^{+}$) infinitely grows as $y^{\delta_{+}(\bar{M}_p)}e^{1/y}$ in this limit. In both cases

$$ ld\,{}_{\infty}\Phi_{\pm,m}[y;a|\pm\vdots\bar{M}_p] \approx \mp y^{-2} \qquad (3.49^{\pm}) $$

and consequently

$$ ld\,{}^{*}\Phi_{\pm,m}[y;a|\pm\vdots\bar{M}_p] \equiv ld\,y - ld\,{}_{\infty}\Phi_{\pm,m}[y;a|\pm\vdots\bar{M}_p] \approx \pm y^{-2} \qquad (3.50^{\pm}) $$

for $0 < y << 1$, where we dropped subscript $\infty$ in the notation of the FF for the reverse RDT:

$$ {}^{*}\Phi_{\pm,m}[y;a|\pm\vdots\bar{M}_p] \equiv y/{}_{\infty}\Phi_{\pm,m}[y;a|\pm\vdots\bar{M}_p]. \qquad (3.51^{\pm}) $$

Note that the last summand in sum ($3.44^{\pm}$) has a simple pole at y=0 so an arbitrary principal solution of RCSLE ($3.37^{\pm}$) near its irregular singular point at y = 0 can be approximated as

$$ {}_{\infty}\Phi_0[y;a;\varepsilon|\pm\vdots\bar{M}_p] \propto y^{\Delta_{\pm}(a;\bar{M}_p)} e^{-1/y} \quad \text{for } y << 1, \qquad (3.52^{\pm}) $$

where $\Delta_{\pm}(a;\bar{M}_p)$ stands for a *finite* power exponent which particular value is non-essential for our discussion. Examination of the quasi-rational function

$$ {}_{\infty}\Phi_0[y;a;\varepsilon|\pm\vdots\bar{M}_{p+1}] = \frac{yW\{{}_{\infty}\Phi_{\pm,m_{p+1}}[y;a|\pm\vdots\bar{M}_p], {}_{\infty}\Phi_0[y;a;\varepsilon|\pm\vdots\bar{M}_p]\}}{{}_{\infty}\Phi_{\pm,m_{p+1}}[y;a|\pm\vdots\bar{M}_p]} \qquad (3.53^{\pm}) $$

$$ = y\,{}_{\infty}\dot{\Phi}_0[y;a;\varepsilon|\pm\vdots\bar{M}_p] - y\,ld\,{}_{\infty}\Phi_{\pm,m_{p+1}}[y;a|\pm\vdots\bar{M}_p]\,{}_{\infty}\Phi_0[y;a;\varepsilon|\pm\vdots\bar{M}_p] $$

representing the RD$\mathfrak{T}$ of the principal solution of RCSLE ($3.37^{\pm}$) near its irregular singular point at y = 0 confirms that it is a principal solution of the transformed RCSLE near the singular point in question. Vice versa the quasi-rational function

$$ \frac{y\,W\{{}^{*}\Phi_{\pm,m_p}[y;a|\pm\vdots\bar{M}_p], {}_{\infty}\Phi_0[y;a;\varepsilon|\pm\vdots\bar{M}_{p+1}]\}}{{}^{*}\Phi_{\pm,m_{p+1}}[y;a|\pm\vdots\bar{M}_p]} \qquad (3.54^{\pm}) $$

$$ = y\,{}_{\infty}\dot{\Phi}_0[y;a;\varepsilon|\pm\vdots\bar{M}_{p+1}] - y\,ld\,{}^{*}\Phi_{\pm,m_{p+1}}[y;a|\pm\vdots\bar{M}_{p+1}]\,{}_{\infty}\Phi_0[y;a;\varepsilon|\pm\vdots\bar{M}_{p+1}] $$

representing the reverse RD$\mathfrak{T}$ of the principal solution ($3.53^{\pm}$) is the principal solution of RCSLE ($3.37^{\pm}$) near its irregular singular point at y = 0.

To study a behavior of Frobenius solutions near a regular singular point of RCSLE ($3.37^{\pm}$) at infinity it is convenient to convert this equation to its 'prime' form [44] using the gauge transformation

$$ {}_{\infty}\Psi[y;a;\varepsilon|\pm\vdots\bar{M}_p] = y^{-1/2}\,{}_{\infty}\Phi[y;a;\varepsilon|\pm\vdots\bar{M}_p] \qquad (3.55^{\pm}) $$

which gives

$$\left\{\frac{d}{dy}y\frac{d}{dy}-y^{-3}+2(a\pm1)y^{-2}+2\frac{d}{dy}\left(y\,ld\,{}_{\infty}\mathcal{W}[y;a\,|\pm\vdots\bar{M}_p]\right)+\varepsilon\,y^{-1}\right\}{}_{\infty}\Psi[y;a;\varepsilon|\pm\vdots\bar{M}_p]=0. \tag{3.56$^{\pm}$}$$

As explained above the main advantage of this representation comes from the fact that the characteristic exponents of two Frobenius solutions of RSLE (3.56$^{\pm}$) near this singular end have opposite signs, with the principal Frobenius solution decaying as $y^{-\sqrt{-\varepsilon}}$ when $y\to\infty$. Again RSLE (3.56$^{\pm}$) is nothing but the 'algebraic' [44] form of the Schrödinger equation with the rational potentials (3.47$^{\pm}$) -- the common feature of RCSLEs with density function (3.2) as far as the given SLE has a regular singular point at infinity. Apparently

$$\begin{aligned}{}_{\infty}\Psi[y;a;\varepsilon|\pm\vdots\bar{M}_{p+1}]&\equiv y^{-1/2}\,{}_{\infty}\Phi[y;a;\varepsilon|\pm\vdots\bar{M}_{p+1}]\\&=\frac{y\,W\{{}_{\infty}\Psi[y;a;\varepsilon_{\pm,m_{p+1}}(a)|\pm\vdots\bar{M}_p],{}_{\infty}\Psi[y;a;\varepsilon|\pm\vdots\bar{M}_p]\}}{{}_{\infty}\Psi[y;a;\varepsilon_{\pm,m_{p+1}}(a)|\pm\vdots\bar{M}_p]}.\end{aligned} \tag{3.57$^{\pm}$}$$

Here we are only interested in cases when the FF appearing in the denominator of PF (3.57$^{\pm}$) is the non-principal Frobenius solution of RSLE (3.56$^{\pm}$) near the singular point at infinity so

$${}_{\infty}\Psi[y;a;\varepsilon|\pm\vdots\bar{M}_{p+1}]\approx-[\sqrt{-\varepsilon}+\sqrt{-\varepsilon_{\pm,m_{p+1}}(a)}]\,y^{-\sqrt{-\varepsilon}}\quad\text{for } y>>1 \tag{3.58$^{\pm}$}$$

if ${}_{\infty}\Psi[y;a;\varepsilon|\pm\vdots\bar{M}_p]$ is an arbitrary principal Frobenius solution of this RSLE near the singular end in question. We thus proved that the RD$\mathfrak{T}$ of any principal Frobenius solution for each of the singular end points is itself the *principal* Frobenius solution of the transformed RSLE near the singular point in question.

Suppose that RSLE (3.56$^{\pm}$) with $\bar{M}_p$ replaced for $\bar{\boldsymbol{M}}^{\pm}_{p+1}$ has an additional eigenfunction ${}_{\infty}\Psi[y;a;\varepsilon^*(a)|\pm\vdots\bar{\boldsymbol{M}}^{\pm}_{p+1}]$ at the energy $\varepsilon^*(a)<0$. Applying the reverse RDT with the FF

$$y^{1/2}/{}_{\infty}\Phi[y;a;\varepsilon_{\pm,m_{p+1}}(a)|\pm\vdots\bar{\boldsymbol{M}}^{\pm}_p]={}_{\infty}\Psi^{-1}[y;a;\varepsilon_{\pm,m_{p+1}}(a)|\pm\vdots\bar{\boldsymbol{M}}^{\pm}_p] \tag{3.59$^{\pm}$}$$

to the new eigenfunction we would come to the solution which obeys the DBC at infinity:

$$\frac{W\{{}_\infty\Psi^{-1}[y;a;\varepsilon_{\pm,m_{p+1}}(a)|\pm\vdots\bar{M}_p^{\pm}],\ {}_\infty\Psi[y;a;\varepsilon^*(a)|\pm\vdots\bar{M}_{p+1}^{\pm}]\}}{{}_\infty\Psi^{-1}[y;a;\varepsilon_{\pm,m_{p+1}}(a)|\pm\vdots\bar{M}_p^{\pm}]} \tag{3.60$^\pm$}$$

$$\approx[\sqrt{-\varepsilon_{\pm,m_{p+1}}(a)}-\sqrt{-\varepsilon^*(a)}]\,y^{-\sqrt{-\varepsilon_{\pm,m_{p+1}}(a)}} \quad \text{for } y >> 1$$

assuming that $\varepsilon^*(a) \neq \varepsilon_{\pm,m_{p+1}}(a)$. On other hand, the quasi-rational function on the left is related to principal solution (3.54$^\pm$) via gauge transformation (3.55$^\pm$) with $\varepsilon = \varepsilon^*(a)$ and therefore the solution in question would obey both DBCs which contradicts the assumption that $\varepsilon^*(a)$ is a new eigenvalue. The only exception corresponds to the case $\varepsilon^*(a) = \varepsilon_{\pm,m_{p+1}}(a)$ when the RDT with FF (3.58$^+$) insert the new bound energy state below the ground energy level of rationally deformed Morse potential (3.47$^+$).

### 3.3. ISOSPECTRAL FAMILY OF RATIONALLY DEFORMED MORSE POTENTIALS WITH A REGULAR SPECTRUM

Let us prove that any set $\bar{M}_p^-$ of seed solutions $\phi_{-,m_k}[y;a]$ $(0 < m_1 < m_k < m_{k+1} \leq p)$ is admissible if the generalized Bessel polynomial $Y_{m_k}^{(-2a)}(y)$ does not have positive zeros so each seed function ${}_\infty\phi_{-,m_k}[y;a]$ preserves its sign on the positive semi-axis. According to (3.34), this is possible iff $m > 2a-1$ for any $m \in \bar{M}_p^-$. In other words we have to prove that polynomial Wronskian (3.43$^-$) does not have positive zeros if this is true for each polynomial $Y_{m_k}^{(-2a)}(y)$. This assertion is obviously trivial for $\tilde{p} = 1$. It also directly follows from the arguments presented in previous subsection that the RDT of ℬRef CSLE (3.1) with the FF $\phi_{-,m_1}[y;a]$ preserves the discrete energy spectrum so the prime RSLE

$$\left\{\frac{d}{dy}y\frac{d}{dy}+y\,{}_\infty I^o[y;a|-\vdots m_1]+(\varepsilon+\tfrac{1}{2})y^{-1}\right\}{}_\infty\Psi[y;a;\varepsilon|-\vdots m_1]=0 \tag{3.61}$$

solved under the DBCs

$$\lim_{y\to 0}{}_\infty\Psi[y;a;\varepsilon_n(a)|-\vdots m_1] = \lim_{y\to\infty}{}_\infty\Psi[y;a;\varepsilon_n(a)|-\vdots m_1] = 0 \tag{3.61'}$$

has exactly N($a$) eigenfunctions

$$_\infty\Psi[y;a;\varepsilon_n(a)|-\vdots\mathfrak{m}_1] \equiv {}_\infty\Psi_{-,n}[y;a|-\vdots\mathfrak{m}_1] = y^{-1/2}\,{}_\infty\Phi_{-,n}[y;a|-\vdots\mathfrak{m}_1] \quad (3.62)$$

at the energies $_\infty\varepsilon_{-,n}(a)$ with n varying from 0 to N($a$) –1. Making use of (3.48‾) with p=1 and m=n, they can be also re-written in the quasi-rational form

$$_\infty\Psi_{-,n}[y;a|-\vdots\mathfrak{m}_1] = {}_\infty\psi_{-,0}[y;a-1]\frac{_\infty\mathfrak{W}[y;a|-\vdots\mathfrak{m}_1,n]}{Y^{(-2a)}_{\mathfrak{m}_1}(y)}. \quad (3.62^\dagger)$$

Keeping in mind that the PF in the right-hand side of the latter expression is proportional to $y^{n-1}$ for y >> 1 one can immediately confirm that eigenfunctions (3.62) vanish in the limit $y \to \infty$ for any $n < a - 1/2$.

Let us now use the mathematical induction to prove that the polynomial $_\infty\mathfrak{W}[y;a|-\vdots\bar{M}^-_{\tilde{p}+1}]$ does not have positive zeros if this assertion holds for the polynomial $_\infty\mathfrak{W}[y;a|-\vdots\bar{M}^-_{\tilde{p}}]$. Again it is suitable to convert RCSLE (3.37⁻) to its prime form

$$\left\{\frac{d}{dy}y\frac{d}{dy} + y\,{}_\infty I^0[y;a|-\vdots\bar{M}^-_{\tilde{p}}] + (\varepsilon + 1/2)y^{-1}\right\}{}_\infty\Psi[y;a;\varepsilon|-\vdots\bar{M}^-_{\tilde{p}}] = 0 \quad (3.63)$$

solved under the DBCs

$$\lim_{y\to 0}\,{}_\infty\Psi[y;a;\varepsilon_n(a)|-\vdots\bar{M}^-_{\tilde{p}}] = \lim_{y\to\infty}\,{}_\infty\Psi[y;a;\varepsilon_n(a)|-\vdots\bar{M}^-_{\tilde{p}}] = 0. \quad (3.63')$$

Making use of (3.48‾) with $p = \tilde{p}$ we can again re-write the eigenfunctions

$$_\infty\Psi_{-,n}[y;a|-\vdots\bar{M}^-_{\tilde{p}}] \equiv {}_\infty\Psi[y;a;\varepsilon_n(a)|-\vdots\bar{M}^-_{\tilde{p}}] = y^{-1/2}\,{}_\infty\Phi_{-,n}[y;a|-\vdots\bar{M}^-_{\tilde{p}}] \quad (3.64)$$

in the quasi-rational form

$$_\infty\Psi[y;a|-\vdots\bar{M}^-_{\tilde{p}+1}] = {}_\infty\psi_{-,0}[y;a-\tilde{p}]\frac{_\infty\mathfrak{W}[y;a|-\vdots\bar{M}^-_{\tilde{p}+1}]}{_\infty\mathfrak{W}[y;a|-\vdots\bar{M}^-_{\tilde{p}}]}. \quad (3.64^\dagger)$$

Examination of q-RS (3.64†) reveals that it vanishes at the origin and therefore represents a principal solution of prime SLE(3.61) near its irregular singular point. Since this solution lies below the lowest eigenvalue it must be nodeless [46] and therefore no Wronskian ${}_{\infty}\mathcal{W}[y; a\,|\,-\vdots\bar{\mathbf{M}}_{p}^{-}]$ has positive zeros.

All the q-RSs

$$ {}_{\infty}\psi_{-,n}[y; a|-\vdots\bar{\mathbf{M}}_{p}^{-}] = {}_{\infty}\psi_{-,0}[y; a-p]\frac{{}_{\infty}\mathcal{W}[y; a\,|\,-\vdots\bar{\mathbf{M}}_{p}^{-}, n]}{{}_{\infty}\mathcal{W}[y; a\,|\,-\vdots\bar{\mathbf{M}}_{p}^{-}]} \tag{3.65} $$

vanish at infinity for n < N($a$) = $\lfloor A \rfloor$ since the power exponent of the PF in the right-hand side of (3.65) is equal to n–p in the limit $y \to \infty$. This confirms that the Direchlet problem for SLE (3.61) has exactly N($a$) eigenfunctions defined via (3.65) with n < N($a$). Since these eigenfunctions must be orthogonal [46] with the unit weight the Wronskians ${}_{\infty}\mathcal{W}[y; a\,|\,-\vdots\bar{\mathbf{M}}_{p}^{-}, n]$ with n varying from 0 to N($a$) –1 are orthogonal with the positive weight

$$ {}_{\infty}\mathrm{W}[y; a\,|\,-\vdots\bar{\mathbf{M}}_{p}^{-}] = \frac{{}_{\infty}\psi^{2}_{-,0}[y; a-p]}{{}_{\infty}\mathcal{W}^{2}[y; a\,|\,-\vdots\bar{\mathbf{M}}_{p}^{-}]} \equiv \frac{{}_{\infty}\phi^{2}_{-,0}[y; a-p+1]}{{}_{\infty}\mathcal{W}^{2}[y; a\,|\,-\vdots\bar{\mathbf{M}}_{p}^{-}]}. \tag{3.66} $$

If the Morse potential has at least 2 energy levels the sequence starts from a polynomial of degree

$$ |\bar{\mathbf{M}}_{p}^{-}| - \tfrac{1}{2}p(p+1) \geq 2p\,, \tag{3.66'} $$

keeping in mind

$$ |\bar{\mathbf{M}}_{p}^{-}| > (2a-1)p + \tfrac{1}{2}p(p+1) > 2p + \tfrac{1}{2}p(p+1) \tag{3.66*} $$

in this case. The finite EOP sequence in question thus starts from a polynomial of at least second degree and therefore [50] does not obey the Bochner theorem [51].

Re-writing (3.44$^{-}$) with $\bar{M}_{p} = \bar{\mathbf{M}}_{p}^{-}$ as

$$ {}_{\infty}I^{0}[y; a\,|\,-\vdots\bar{\mathbf{M}}_{p}^{-}] = {}_{\infty}I^{0}[y; a-p] + \frac{2}{y}\frac{d}{dy}\left(y\,ld\;{}_{\infty}\mathcal{W}[y; a\,|\,-\vdots\bar{\mathbf{M}}_{p}^{-}]\right) \tag{3.67} $$

we can then explicitly express corresponding Liouville potential (3.47$^{-}$) in terms of the admissible Wronskian ${}_{\infty}\mathcal{W}[y; a\,|\,-\vdots\bar{\mathbf{M}}_{p}^{-}]$ as follows

$$ {}_{\infty}V[y; a\,|\,-\vdots\bar{\mathbf{M}}_{p}^{-}] = {}_{\infty}V[y; a-p] - 2y\frac{d}{dy}\left(y\,ld\;{}_{\infty}\mathcal{W}[y; a\,|\,-\vdots\bar{\mathbf{M}}_{p}^{-}]\right). \tag{3.67†} $$

As mentioned in previous subsection this net of isospectral rational potentials starts from potential function (2.15) in [10] with $A = a - ½, B = 1$, after the latter is expressed in terms of the variable $y=e^x$.

### 3.4. SUBNET OF RATIONALLY DEFORMED MORSE POTENTIALS QUANTIZED VIA WRONSKIANS OF R-BESSEL POLYNOMIALS

Another family of solvable RDC𝒮s of CSLE (3.1) can be constructed using juxtaposed pairs of eigenfunctions ${}_\infty\phi_{-,n_k}[y;a]$, ${}_\infty\phi_{-,n_k+1}[y;a]$ ($0 < n_k < n_{k+1} - 1 < N(a)$ for $k = 1, \ldots, J$). The simplest double-step representative of this finite family of rationally deformed Morse potentials with $n_1$=1, J=2 was constructed by Bagrov and Samsonov [40, 48] in the late nineties based on the conventional ℒRef representation of the Schrödinger equation with the Morse potential. The extensions of their works to an arbitrary number of juxtaposed pairs of eigenfunctions in both ℒRef and ℬRef representations were performed more recently in [11] and [19] accordingly.

For any TFI RCSLE from Group A one can by-pass an analysis of the pre-requisites for the Krein-Adler theorem [52, 53] by taking advantage of the fact that the Wronskians of eigenfunctions are composed of weighted orthogonal polynomials with the common degree-independent weight and therefore the numbers of their positive zeros are controlled by the general Conjectures proven in [54] for Wronskians of positive definite orthogonal polynomials. In particular we conclude that any Wronskian formed by juxtaposed pairs of R-Bessel polynomials of non-zero degrees may not have positive zeros.

Let $\bar{\mathbf{N}}_{2J}$ be a set of R-Bessel polynomials of degrees

$$\bar{\mathbf{N}}_{2J} = \bar{M}(\bar{\Delta}'_{L\to 1}) = n_1 : n_1 + 2j_1 - 1, n_{2j_1+1} : n_{2j_1+1} + 2j_2 - 1, \ldots, n_{2J-2j_L+1} : n_{2J}$$

$$(n_1 > 0,\ n_{2J} < N) \qquad (3.68)$$

with even

$$\delta'_l = 2j_l \quad (l = 1, \ldots, L). \qquad (3.68')$$

Examination of the q-RS functions

$$ {}_{\infty}\Psi_{-,n}[y;a|-\vdots\bar{\mathbf{N}}_{2J}] = y^{-1/2}\,{}_{\infty}\Phi_{-,n}[y;a|-\vdots\bar{\mathbf{N}}_{2J}] = {}_{\infty}\psi_{-,0}[y;a-2J]\frac{{}_{\infty}\mathcal{W}[y;a\,|-\vdots\bar{\mathbf{N}}_{2J},n]}{{}_{\infty}\mathcal{W}[y;a\,|-\vdots\bar{\mathbf{N}}_{2J}]} $$

$$ (n \notin \bar{\mathbf{N}}_{2J}) \qquad (3.69) $$

shows that they all represent principal solutions near the irregular singular point of the prime RSLE

$$ \left\{\frac{d}{dy}y\frac{d}{dy} + y\,{}_{\infty}I^{o}[y;a|-\vdots\bar{\mathbf{N}}_{2J}] + (\varepsilon + 1/2)y^{-1}\right\}{}_{\infty}\Psi[y;a;\varepsilon|-\vdots\bar{\mathbf{N}}_{2J}] = 0 \qquad (3.70) $$

assuming again that the latter equation is solved under the DBCs

$$ \lim_{y\to 0}\,{}_{\infty}\Psi[y;a;\varepsilon_n|-\vdots\bar{\mathbf{N}}_{2J}] = \lim_{y\to\infty}\,{}_{\infty}\Psi[y;a;\varepsilon_n|-\vdots\bar{\mathbf{N}}_{2J}] = 0\,. \qquad (3.70') $$

Note that the PF in the right-hand side of (3.69) is proportional to $y^{n-2J}$ for $y >> 1$ so each solution with $n \notin \bar{\mathbf{N}}_{2J} < N(a)$ represents an eigenfunction of RSLE (3.70).

Again these eigenfunctions must be orthogonal with the weight $y^{-1}$ and therefore $N(a)-2J$ Wronskians ${}_{\infty}\mathcal{W}[y;a\,|-\vdots\bar{\mathbf{N}}_{2J},n]$ with $n \notin \bar{\mathbf{N}}_{2J} < N(a)$ form a polynomial set orthogonal with the positive weight

$$ {}_{\infty}\mathrm{W}[y;a\,|-\vdots\bar{\mathbf{N}}_{2J}] = \frac{{}_{\infty}\psi^{2}_{-,0}[y;a-2J]}{{}_{\infty}\mathcal{W}^{2}[y;a\,|-\vdots\bar{\mathbf{N}}_{2J}]} \equiv \frac{{}_{\infty}\phi^{2}_{-,0}[y;a-2J+1]}{{}_{\infty}\mathcal{W}^{2}[y;a\,|-\vdots\bar{\mathbf{N}}_{2J}]}\,. \qquad (3.71) $$

If sequence (3.68) starts from $n_1 = 1$ then the finite EOP sequence in question lacks the first-degree polynomial. Otherwise it always starts from a polynomial of non-zero degree

$$ |\bar{\mathbf{N}}_{2J}| - J(2J+1) > (n_1 - 1)(\eth'_1 - 1) \geq 1\,. \qquad (3.72) $$

In both cases the pre-requisites of the Bochner theorem are invalid as expected [50].

The Liouville potentials in question can be thus expressed in terms of the admissible Wronskians ${}_{\infty}\mathcal{W}[y;a\,|-\vdots\bar{\mathbf{N}}_{2J}]$ as follows

$$ {}_{\infty}V[y;a\,|-\vdots\bar{\mathbf{N}}_{2J}] = {}_{\infty}V[y;a-2J] - 2y\frac{d}{dy}\left(y\,ld\,{}_{\infty}\mathcal{W}[y;a\,|-\vdots\bar{\mathbf{N}}_{2J}]\right). \qquad (3.73) $$

We refer the reader to Conjectures in [54] to verify that the number of zeros of each Wronskian in the constructed orthogonal polynomial set changes exactly by 1 even if a jump in the polynomial degree is larger than 1. However even if we take advantage of these elegant results we still need to prove that there are no additional eigenfunctions with a number of nodes larger than $N(a) - 2J - 1$. In contrast with the analysis presented in the previous Section, this proof is complicated by the fact that the RDT at each odd step results in a non-solvable RSLE with singularities on the positive semi-axis. Luckily we deal with the TFI CSLE so its RDC$\mathfrak{I}$ using juxtaposed pairs of eigenfunctions can be alternatively obtained via *sequential* RDTs with seed solutions from the second sequence +,m [19, 11]. Namely, as already mentioned in the end of Section 2 the conjugated partition

$$\bar{\mathbf{M}}^{+}_{|\bar{\delta}_{1\to L}|} = \bar{M}(\bar{\Delta}_{1\to L}) \tag{3.74}$$

is formed by alternating even and odd integers starting from an even integer $\bar{\delta}'_1$. The reverse is also true: if the partition

$$\bar{\mathbf{M}}^{+}_{p} = \bar{M}({}^{p}\bar{\delta}_{1\to L_p}; {}^{p}\bar{\delta}'_{1\to L_p}) \quad \bar{\mathbf{M}}^{+}_{p} = \bar{M}({}^{p}\bar{\delta}_{1\to L_p}; {}^{p}\bar{\delta}'_{1\to L_p}) \tag{3.75}$$

is composed of alternating even and odd integers starting from an even integer $\bar{\delta}'_1$ then each segment of the conjugated partition

$${}^{p}\bar{\mathbf{N}}_{2J_p} = \bar{M}({}^{p}\bar{\delta}'_{L_p\to 1}; {}^{p}\bar{\delta}_{L_p\to 1}) \tag{3.75$^{\dagger}$}$$

must have an even length, with the largest element

$$\mathbf{m}^{+}_{|{}^{p}\bar{\delta}'_{1\to L_p}|} = |{}^{p}\bar{\Delta}_{1\to L_p}| - 1 = \mathbf{m}_{|{}^{p}\bar{\delta}_{L_p\to 1}|} \in {}^{p}\bar{\mathbf{N}}_{2J_p}, \tag{3.76}$$

where ${}^{p}\bar{\Delta}_{1\to L_p} \equiv {}^{p}\bar{\delta}_{1\to L_p}; {}^{p}\bar{\delta}'_{1\to L_p}$.

Making use of (3.5$^{\pm}$) one can verify that quasi-rational functions (2.27$^{\pm}$) can be decomposed as

$${}_{\infty}\chi_{\mp N}[y;a] = y^{\frac{1}{2}N(N-1)\mp N\delta}[\xi]_{\infty}\phi^{N}_{\pm,0}[y;a^{(\delta)}] \tag{3.77$^{\pm}$}$$

and therefore the denominators of the fractions in equivalence relations (2.26) take form

$$y^{-\frac{1}{2}|\bar{\delta}_L|(|\bar{\delta}_L|-1)} \prod_{l=1}^{L} {}_{\infty}\chi_{-\delta_l}[y; a^{(|\bar{\Delta}'_{l\to 1}|-\delta_l)}] = y^{\Sigma_L}\, {}_{\infty}\phi_{-,0}^{|\bar{\delta}_L|}[y; a] \tag{3.78}$$

$$y^{-\frac{1}{2}|\bar{\delta}'_L|(|\bar{\delta}'_L|-1)} \prod_{l=1}^{L} {}_{\infty}\chi_{\delta'_l}[y; a^{(|\bar{\Delta}'_{l-1\to 1}|+\delta'_l)}] = y^{\Sigma_L}\, {}_{\infty}\phi_{-,0}^{|\bar{\delta}'_L|}[y; a^{(|\bar{\Delta}_{1\to L}|)}] \tag{3.78'}$$

accordingly, where $|\bar{\Delta}_{1\to L}| = |\bar{\Delta}'_{L\to 1}|$

$$\Sigma_L \equiv \sum_{l=1}^{L} \delta'_l (\delta_l + \sum_{\underset{\sim}{l}=l+1}^{L} \delta_{\underset{\sim}{l}}) = \sum_{l=1}^{L} \delta_l (\delta'_l + \sum_{\underset{\sim}{l}=1}^{l-1} \delta'_{\underset{\sim}{l}}). \tag{3.79}$$

We thus come to the following equivalence theorem for the Wronskians of generalized Bessel polynomials

$${}_{\infty}\hat{\mathcal{W}}[y; a\,|+\vdots\bar{M}(\bar{\Delta}_{1\to L})] = {}_{\infty}\hat{\mathcal{W}}[y; a^{(|\bar{\Delta}_{1\to L}|)}\,|-\vdots\bar{M}(\bar{\Delta}'_{L\to 1})]. \tag{3.80}$$

Note that decomposition (3.77$^{\pm}$) holds for any TFI CSLE of Group A provided that we replace $y^2$ for the leading coefficient ${}_{\iota}\sigma[y]$ of the corresponding counter-parts of differential eigenequations (3.14$^{\pm}$). This brings us to the equivalence relations for polynomial Wronskians discovered by Odake and Sasaki [19] in their pioneering analysis of TSI potentials from Group A.

If $a > \frac{1}{2}$ then, according to (3.76), the largest element of the partition ${}^{p}\bar{\mathbf{N}}_{2J_p}$ is smaller than $a + |{}^{p}\bar{\Delta}_{1\to L_p}| - \frac{1}{2}$ and therefore the Wronskian in the right-hand side of (3.80) with $\bar{\Delta}'_{L\to 1}$ replaced for ${}^{p}\bar{\Delta}'_{L_p\to 1}$ is formed by juxtaposed pairs of R-Bessel polynomials. This confirms that none of the polynomial Wronskians ${}_{\infty}\hat{\mathcal{W}}[y; a\,|+\vdots\bar{\mathbf{M}}_p^+]$ has zeros on the positive semi-axis and therefore each partition $\bar{\mathbf{M}}_p^+$ specifies an admissible sequence of seed solutions ${}_{\infty}\phi_{+,\mathbf{m}_k}[y; a]$ ($\mathbf{m}_k \in \bar{\mathbf{M}}_p^+$ for k=1,…,p). Based on the arguments presented in subsection 3.2 we thus assert that the RDTs in question may insert only one bound energy level at the energy ${}_{\infty}\varepsilon_{+,\mathbf{m}_{p+1}}(a)$ which by definition lies below the ground energy level ${}_{\infty}\varepsilon_{+,\mathbf{m}_p}(a)$ of the Liouville potential ${}_{\infty}V[a\,|\,\bar{\mathbf{M}}_p^+]$. On other hand all the existent energy levels remain unchanged.

As the simplest example we can cite the partition

$$1,2,\ldots,2J=\bar{M}(2J,1)=\bar{M}^{\dagger}(1,2J) \text{ for } 2J \le \lfloor A \rfloor \tag{3.81}$$

As a direct consequence of the equivalence theorem we find that

$$\hat{Y}_{2J}^{(2a-2J-1,-2)}(y)={}_{\infty}\hat{\mathcal{W}}[y;a\,|\,-\vdots 1:2J] \quad (2J \le \lfloor A \rfloor), \tag{3.82}$$

where the Wronskian on the right is formed by 2J sequential R-Bessel polynomials of non-zero degrees smaller than A and therefore may not have positive zeros for $a>-\frac{1}{2}$. As initially proven in [7] and then illuminated in more details in [8] using the so-called ‘Kienast-Lawton-Hahn’s theorem’ [55-57] the latter assertion holds for any positive J despite the fact that the seed functions ${}_{\infty}\psi_{-,m}[y;a^{(2J+1)}]$ have nodes on the positive semi-axis for

$$a^{(2J+1)}+\frac{1}{2}<m<2A. \tag{3.83}$$

Indeed, representing (3.23) as

$$Y_m^{(2a,-2)}(y)\equiv Y_m^{(2a)}(-y/2)=m!\left(-y/2\right)^m L_m^{(-2a-2m-1)}\left(-2/y\right) \tag{3.84}$$

shows that the absolute value of the negative m-dependent Laguerre index

$$\alpha_m=-2a-2m-1<0 \tag{3.85}$$

is larger than the polynomial degree and therefore the polynomial in question may not have zeros at negative values of its argument.

## 3.5. Isospectral rational extensions of Krein-Adler SUSY partners of Morse potential

Since any RDCT of the Morse potentials using pairs of juxtaposed eigenfunctions $\bar{\mathbf{N}}_{2J}$ keeps unchanged the ground-energy level a set of seed functions ${}_{\infty}\phi_{+,\mathbf{m}}[y;a]$ is admissible iff all $\mathbf{m}\in\bar{\mathbf{N}}_{2J},\bar{\mathbf{M}}_p^-$, where $\bar{\mathbf{M}}_p^-$ is an admissible set of seed polynomials specified in subsection 3.3. We can then use the same arguments as in subsection 3.3 to prove that any Liouville potential

$$ {}_{\infty}V[y;a\,|\,-\vdots\bar{\mathbf{N}}_{2J},\bar{\mathbf{M}}_{p}^{-}] = {}_{\infty}V[y;a-2J-p] - 2y\frac{d}{dy}\left(y\,ld\;{}_{\infty}\mathcal{W}[y;a\,|\,-\vdots\bar{\mathbf{N}}_{2J},\bar{\mathbf{M}}_{p}^{-}]\right) \tag{3.85} $$

has exactly the same discrete energy spectrum as rationally deformed Morse potential (3.73) constructed by means of juxtaposed pairs of R-Bessel polynomials of non-zero degrees. Its eigenfunctions expressed in terms of the variable $y=e^{x}$ can be represented as

$$ {}_{\infty}\Psi_{-,n}[y;a|-\vdots\bar{\mathbf{N}}_{2J},\bar{\mathbf{M}}_{p}^{-}] = {}_{\infty}\psi_{-,0}[y;a-2J-p]\,\frac{{}_{\infty}\mathcal{W}[y;a\,|\,-\vdots\bar{\mathbf{N}}_{2J},\bar{\mathbf{M}}_{p}^{-},n]}{{}_{\infty}\mathcal{W}[y;a\,|\,-\vdots\bar{\mathbf{N}}_{2J},\bar{\mathbf{M}}_{p}^{-}]} \tag{3.86} $$

for $n \notin \bar{\mathbf{N}}_{2J} < N(a)$,

keeping in mind that the corresponding prime RSLE is nothing but the Schrödinger equation re-written in its algebraic form.

Each polynomial sequence ${}_{\infty}\mathcal{W}[y;a\,|\,-\vdots\bar{\mathbf{N}}_{2J},\bar{\mathbf{M}}_{p}^{-},n]$ for $n \notin \bar{\mathbf{N}}_{2J} < N(a)$ is X-orthogonal:

$$ \int_{0}^{\infty} {}_{\infty}\mathcal{W}[y;a\,|\,-\vdots\bar{\mathbf{N}}_{2J},\bar{\mathbf{M}}_{p}^{-},n]\,{}_{\infty}\mathcal{W}[y;a\,|\,-\vdots\bar{\mathbf{N}}_{2J},\bar{\mathbf{M}}_{p}^{-},n']\,{}_{\infty}W[y;a\,|\,-\vdots\bar{\mathbf{N}}_{2J},\bar{\mathbf{M}}_{p}^{-}]\,dy = 0 \tag{3.87} $$

for $n < n' < N(a)$

with the weight

$$ {}_{\infty}W[y;a\,|\,-\vdots\bar{\mathbf{N}}_{2J},\bar{\mathbf{M}}_{p}^{-}] = \frac{{}_{\infty}\psi_{-,0}^{2}[y;a-2J-p]}{{}_{\infty}\mathcal{W}^{2}[y;a\,|\,-\vdots\bar{\mathbf{N}}_{2J},\bar{\mathbf{M}}_{p}^{-}]} \equiv \frac{{}_{\infty}\phi_{-,0}^{2}[y;a-2J-p+1]}{y\,{}_{\infty}\mathcal{W}^{2}[y;a\,|\,-\vdots\bar{\mathbf{N}}_{2J},\bar{\mathbf{M}}_{p}^{-}]}\,. \tag{3.88} $$

## 4. CONCLUSIONS

The presented analysis illuminates the non-conventional approach [19] to the family of rationally deformed Morse potentials using seed solutions expressed in terms of Wronskians of generalized Bessel polynomials in the variable $y = e^{x}$. As a new achievement compared with Odake and Saski's [19] study on RDC$\mathfrak{J}$s of the Morse potential (see also [11] where a similar analysis was performed within the conventional $\mathcal{L}$Ref framework) we constructed a new RDC net of *isospectral* potentials by expressing them in terms of the logarithmic derivative of Wronskians of generalized Bessel polynomials with no positive zeros. The constructed isospectral family of rationally deformed Morse potentials represents a natural extension of the isospectral RD$\mathcal{T}$s of the Morse potential discovered by Quesne [10].

An important element of our analysis often overlooked in the literature is the proof that the sequential RDTs in question do not insert new bound energy states. The widespread argumentation in support of this (usually taken-for-granted) presumption is based on the speculation that the theorems of the *regular* Sturm-Liouville theory [58] are automatically applied to singular SLEs. We can refer the reader to the scrupolous analysis performed in [46] for SLEs solved under the DBCs as an illustration that this is by no means a trivial issue.

To be able to prove the aforementioned assertion we converted the given RCSLE to its prime form such that the characteristic exponents of Frobenius solutions for the regular singular point at $\infty$ have opposite signs and therefore the principal Frobenius solution near this singular end is unambiguously selected by the corresponding DBC. (In the particular case under consideration the prime RSLE accidently coincides with the Schrödinger equation re-written in the 'algebraic' [45] form but this is not true in general.) Re-formulating the given spectral problem in such a very specific way allowed us to take advantage of powerful theorems proven in [46] for zeros of principal solutions of SLEs solved under the DBCs at singular ends. We [45] also used this simplified version of the conventional spectral theory to prove that any RD$\mathcal{T}$ of a principal (non-principal) Frobenius solution near the regular singular point at $\infty$ is itself a principal (non-principal) Frobenius solution of the transformed RSLE. This assertion plays a crucial role in our proof of the exact solvability of the constructed DC net of isospectral rational potentials.

It is commonly presumed that the Krein-Adler theorem [52, 53] is applied to an arbitrary potential regardless its behavior near the singular end points. In [45] we examined this presumption more carefully for the Dirichlet problems of our interest again taking advantage of the theorems proven in [45] for zeros of juxtaposed eigenfunctions. However one can by-pass this analysis for any TFI RSLE from Group A keeping in mind that the Wronskians in questions are formed by orthogonal polynomials with degree independent indexes and therefore the numbers of their positive zeros are controlled by the general Conjectures proven in [53]. In particular this implies that any Wronskian formed by juxtaposed pairs of R-Bessel polynomials of non-zero degrees may not have positive zeros.

## ACKNOWLEDGEMENTS

I am grateful to A. D. Alhaidari for bringing my attention to the alternative representation of eigenfunctions of the Schrödinger equation with the Morse potential in terms of R-Bessel polynomials with *degree-independent* indexes. This alteration helped me to fully comprehend Odake and Sasaki's suggestion to place the Morse oscillator into Group A of rational TSI potentials.